\documentstyle[aps,epsfig]{revtex}
\begin{document}

\twocolumn[
 \hsize\textwidth\columnwidth\hsize
 \csname@twocolumnfalse\endcsname

\draft
\title{The charge and low-frequency response of normal-superconducting 
heterostructures}
\author{S. Pilgram$^1$, H. Schomerus$^2$,  A. M. Martin$^3$, M. B\"uttiker$^1$}

\address{$^1$Dept. Phys. Th\'eorique, Universit\'e de Gen\`eve,
24, quai Ernest-Ansermet, 1211 Gen\`eve 4, Switzerland}

\address{$^2$Max-Planck-Institut f\"ur Physik komplexer Systeme,
N\"othnitzer Str. 38, 01187 Dresden, Germany}

\address{$^3$School of Physics and Astronomy, University of Nottingham, Nottingham, 
NG7 2RD,  U.K.}

\maketitle

\begin{abstract}
The charge distribution is a basic aspect of electrical transport. In 
this work we investigate the self-consistent charge response of 
normal-superconducting heterostructures. Of interest is the variation 
of the charge density due 
to voltage changes at contacts 
and due to changes in the potential. 
We present response functions in terms of functional 
derivatives of the scattering matrix. 
We discuss corrections to the Lindhard function due to the proximity 
of the superconductor. We use these results 
to find the dynamic conductance matrix to lowest order in frequency. 
We illustrate similarities
and differences between normal systems and heterostructures
for specific examples like a ballistic wire, a resonator, and a quantum 
point contact. 
\end{abstract}


]

\section{Introduction}

During the past decade mesoscopic systems consisting of 
both normal and superconducting parts have attracted 
considerable attention. Microscopically, the interesting physics 
stems from Andreev reflection. An incident particle is reflected as
a hole and a Cooper pair is generated in the superconductor. This results 
in an effective charge transfer of $2e$ and correlations between 
Andreev reflected electron hole pairs (the proximity effect).  
These effects have been investigated in many experimental  
and theoretical works {\onlinecite{tinkh,super}} 
focusing mainly on the stationary
transport regime (dc-conductance) \cite{vanWees1,Pannetier1} 
and the low-frequency noise 
(shot noise) \cite{Sanquer1,Schoenenberger1}. The ac-regime has
attracted much less attention \cite{Takayanagi1,Takayanagi2,schoelkopf,Wang1}.

In an Andreev process the electron and hole parts of the wave function
contribute with opposite charge. It is therefore interesting to
investigate the low-frequency ac-transport of NS-systems, since
this problem requires a electrically self-consistent
discussion of the charge distribution in the sample.
This self-consistency is of importance not
only for ac-transport but also for the discussion of charge fluctuations 
and the non-linear transport regime \cite{Andrew1}. 

In this work we have in mind the interplay of two
main properties of hybrid structures: On one hand
raising or lowering the voltage 
at a normal contact of the sample will not inject an additional charge 
into regions where the wave functions contain electron and
hole amplitudes of equal magnitude.
This is in strong contrast to a purely normal conductor!
On the other hand screening is a property not only 
of the states at the Fermi surface but of the entire electron gas. 
Thus the ability of a hybrid structure to screen an additional 
charge is essentially the same as that of a normal conductor.

Our results show two main differences compared to purely normal systems:
first, the coupling of carriers with opposite charge reduces the 
interaction with nearby gates. Second, 
Andreev reflection increases the dwell time
inside the structure and this affects the ac-response. 

The paper is organized as follows: 
In section \ref{chde}, we derive an expression of 
the charge density in terms of functional derivatives 
of the scattering matrix. We next discuss the charge 
density response to external and internal 
potential perturbations. 
In section \ref{linres} we use these results 
to formulate a self-consistent
theory of low-frequency ac-response. 
To illustrate our results we consider in section \ref{examples} 
a number of examples: a ballistic wire, a resonant structure 
and the quantum point contact connected to a superconductor. 

\section{Charge Density}
\label{chde}

\begin{figure}[htb]
\begin{center}
\leavevmode
\psfig{file=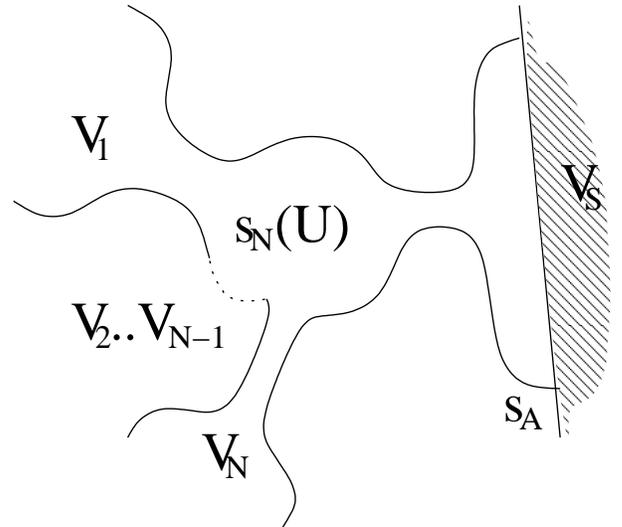,width=8cm}
\caption{A mesoscopic scattering region is attached to $N$ normal
reservoirs and one superconducting reservoir. It is described by
two scattering matrices $s_N$ describing the normal conducting
region and $s_A$ describing the Andreev reflection. Each terminal
has its own bias voltage $V_{\alpha}$. The electrostatic
potential $U$ inside the scatterer is calculated self-consistently.
}
\label{Scattering Problem}
\end{center}
\end{figure}

In this section we derive the general expression for the
charge density $\langle \hat{\rho}(r) \rangle$
in the scattering problem sketched in Fig. \ref{Scattering Problem}.
A scattering region is attached to $N$ normal leads and one superconducting
lead. Every normal lead is characterized by its applied voltage
$V_{\alpha}$, the superconducting lead by its pair potential $\Delta$ and
the bias $V_S$. For all the calculations we may choose $V_S=0$. 
The fact that we only allow for one superconducting
terminal excludes all time-dependent Josephson-like effects.
For an introduction to the applied formalism 
we refer the reader to Ref. \cite{Datta1}. 
The whole system is described by its scattering matrix
\begin{equation}
\label{Scattering Matrix}
s_{\alpha\beta} = \left(\begin{array}{cc}
s_{\alpha\beta}^{pp} & s_{\alpha\beta}^{ph}\\
s_{\alpha\beta}^{hp} & s_{\alpha\beta}^{hh}\\
\end{array}\right).
\end{equation}
The element $s_{\alpha\beta}^{hp}$ for example is the current
amplitude of a hole that leaves through lead $\alpha$ 
and has entered with unit current amplitude 
as a particle through lead $\beta$. 
We represent each scattering channel
by its own lead to save two indices. 

It is conceptually useful \cite{Beenakker1} to imagine
the scattering matrix $s$ being assembled from a part that describes
the reflection and transmission in the normal region $s_N^{p/h}$ and from a
part that describes the Andreev processes at the interface between
normal metal and superconductor $s_A$. Only this second matrix leads to coupling
between the particle and hole scattering states.
However, the following derivations do not depend on this assumption.
For energies $E$ smaller than the
gap energy $|\Delta|$ only reflection takes place at the interface. The total
scattering matrix $s$ then has the dimension $(2N_N)^2$ where $N_N = \sum N_{\alpha}$ is the
total number of channels leading to normal reservoirs.
Above the gap
we must also include transmission processes, and therefore
the dimension changes to $(2N+2N_S)^2$. The superconducting terminal
adds $N_S$ more channels.

For the following it is helpful to introduce local partial 
densities of states (LPDOS) \cite{Gramespacher1}
\begin{equation}
\label{Partial Densities}
\nu(\alpha_{\nu},r_{\kappa},\beta_{\mu}) = -\frac{1}{2\pi q^{\kappa}}\text{Im} 
\left[(s_{\alpha\beta}^{\nu\mu})^*\frac{\delta s_{\alpha\beta}^{\nu\mu}}{\delta U^{\kappa}(r)}\right].
\end{equation}
This expression is valid for one channel per lead.
A true multichannel expression would include
a trace over the channels.
The value $\nu(1^h,r^p,2^p)$ for example describes the density of
particles at location $r$ that entered as particles through
contact $2$ and leave as holes through lead $1$. 
In definition (\ref{Partial Densities}) we denote the quasi-particle charge by $q^{p/h} = \pm q$.
The LPDOS must be calculated as functional derivatives of the
scattering matrix with respect to the electrostatic potential $U$.
To gain the information about particles and holes separately \cite{Gramespacher1}
we artificially split up the electrostatic potential $U$ in a part that
acts on particles $U^p$ and another that addresses holes $U^h$. The Bogoliubov-de Gennes Hamiltonian
then takes the form
\begin{equation}
\label{Bogoliubov-de Gennes}
\hat{H} = \left(\begin{array}{l}
H_0+qU^p(x)-E_F \quad \Delta \\
\quad\\
\Delta^* \quad -(H_0+qU^h(x)-E_F)^*\\
\end{array}\right).
\end{equation}
These equations have to be solved including a small variation
of the electrostatic potentials $U^p$ and $U^h$ in order to
get the scattering matrix and its functional derivatives.

The above defined LPDOS are not
independent. On one hand, the LPDOS 
obey reciprocity relations. 
This has been investigated in reference \cite{Gramespacher1}. 
On the other hand
the particle-hole symmetry of the Bogoliubov-de Gennes equation implies
\begin{equation}
\label{Particle-Hole Symmetry}
\nu(\alpha_{\nu},r_{\kappa},\beta_{\mu},E) = \nu(\alpha_{\bar{\nu}},r_{\bar{\kappa}},\beta_{\bar{\mu}},-E).
\end{equation}
The bar denotes the opposite ($p/h = \bar{h}/\bar{p}$). Both symmetries can be used to
reduce the expense of the calculation. 

The charge density inside the normal-superconducting heterostructure can be entirely expressed
by the LPDOS (and therefore by the scattering matrix) 
and the occupation factors of the attached
reservoirs
\begin{equation}
\label{Charge Density}
\begin{array}{l}
\langle \hat{\rho}(r) \rangle = q\int_{-\infty}^{\infty} dE \sum_{\alpha\beta\mu\nu}\\
\quad\\
\quad \left(
f_{\beta}^{\mu}(E) \nu(\alpha_{\nu},r_p,\beta_{\mu}) + (1-f_{\beta}^{\mu}(E)) \nu(\alpha_{\nu},r_h,\beta_{\mu})
\right).
\end{array}
\end{equation}
The occupation factors include the bias voltage of the normal terminals
$f_{\beta}^{p/h}(E) = f(E-q^{p/h}V_{\beta})$. 
Here $f$ is the Fermi function. Note that the occupation factors
vary in opposite
directions for particles and holes. In Eq. (\ref{Charge Density}) 
we have double counted the particle-hole excitations 
and hence drop a factor of two for spin degeneracy. 
The derivation of this result is
outlined in appendix \ref{Anhang A}.

\section{Charge Response and Gauge Invariance}
\label{chres} 

Given formula (\ref{Charge Density}) we are now in a position to calculate the
charge density response $\delta \rho(r)$ to both internal and external potential
variations
\begin{equation}
\label{Charge Response}
\delta \rho(r) = \sum_{\beta=S,1..N} 
\frac{\partial \rho(r)}{\partial V_{\beta}}\delta V_{\beta}
- \int dr' \tilde{\Pi}(r,r')\delta U(r').
\end{equation}
The first contribution is the bare charge injected from the leads due to
the shift of the occupation factors, and is proportional to the
injectivities $\partial \rho(r)/\partial V_\beta$. The second
contribution arises from the change of the internal potential due to
screening (the potential itself will be determined in the following
subsection), and involves the Lindhard function
$\tilde\Pi(r,r')=-\partial \rho(r)/\partial U(r')$.

The injectivity from the normal leads can be calculated straightforwardly from
the charge density (\ref{Charge Density})
\begin{equation}
\label{Normal Injectivity}
\frac{\partial\rho(r)}{\partial V_{\beta}} = \int dE 
\left(-\frac{\partial f}{\partial E}\right) \sum_{\alpha\nu\mu\kappa}
q^{\kappa}q^{\mu} \nu(\alpha_{\nu},r_{\kappa},\beta_{\mu})
\end{equation}
and depends at low temperatures as expected only on properties at the Fermi energy.
The other quantities contained in the balance equation (\ref{Charge Response}) need
a more careful analysis. Its technical details are explained in appendix
\ref{Anhang B}.

The procedure of calculating the non-local Lindhard function $\tilde{\Pi}(r,r')$ 
leads to second order functional
derivatives that cannot be simplified further.
However, if we assume the
Lindhard function to be local, $\tilde{\Pi}(r,r') = \delta(r-r')\Pi(r)$,
we can express it by the above calculated LPDOS (\ref{Partial Densities}).
This assumption is correct if the
electrostatic potential varies only slowly on the scale of the
Fermi wavelength $\lambda_F$.
To express the result in a transparent way we split up the energy $E$ in the
same way as the potential $U(r)$. We introduce the quantities $E^{p/h}$ that
denote the energy dependence of the scattering matrices for particles and holes $s_N^{p/h}(E^{p/h})$.
Furthermore we need a critical energy $E_c$ with the property
$|\Delta| \ll E_c \ll E_F$. Such an energy is useful, since particles and holes with energies outside the
range $[-E_c,E_c]$ have a negligible Andreev reflection probability
that decays as $\sim |\Delta|^2/E^2$. In a short calculation given 
in appendix \ref{Anhang B} we
find for the local Lindhard function
\begin{equation}
\label{Lindhard Function}
\begin{array}{ll}
\Pi(r) = & q^2 \sum_{\alpha\beta}\left[\nu(\alpha_{p},r_{p},\beta_{p},-E_c) +
                                          \nu(\alpha_{h},r_{h},\beta_{h},+E_c)\right]\\
\quad\\
 & + q^2 \int_{-E_c}^{+E_c} dE \sum_{\alpha\beta\mu\nu}
         \left.\left(\frac{\partial}{\partial E^p} - \frac{\partial}{\partial E^h}\right)\right|_{E^p,E^h=E}\\
\quad\\
 & \left[f(E) \nu(\alpha_{\nu},r_p ,\beta_{\mu}) + f(-E) \nu(\alpha_{\nu},r_h ,\beta_{\mu})\right].\\
\end{array}
\end{equation}
If the local electrostatic potential $U(r)$ fulfills the condition
$E_F-U(r)\gg|\Delta|$ the first line
of Eq. (\ref{Lindhard Function}) dominates. Therein we can replace
the sum over the LPDOS by the local density of states at $E=0$
for the equivalent normal conducting structure. 
This is the expression for the Thomas-Fermi screening in a purely normal
sample. The screening properties are not affected by the presence of the
superconductor.
Our result becomes more interesting for $E_F-U(r)\sim |\Delta|$ when the
second and third term of Eq. (\ref{Lindhard Function})
contribute significantly. 
We later illustrate this case 
with a specific example.

A simple argument allows us to get the injectivity from the superconducting terminal
$\partial \rho(r)/ \partial V_S$ without any further calculation. The 
Bogoliubov-de Gennes
equations are gauge invariant, a simultaneous change of all 
external and internal potentials by
the same amount will not lead to any charge inside the system. Setting the left side
of Eq. (\ref{Charge Response}) to zero gives therefore
\begin{equation}
\label{Gauge Invariance}
\Pi(r) = \sum_{\beta=S,1..N} \frac{\partial \rho(r)}{\partial V_{\beta}}.
\end{equation}
Since $\Pi(r)$ and the injectivities of all normal contacts 
are known 
we can use this relation to find the injectivity 
$\partial \rho(r)/ \partial V_S$ of the superconducting contact. 

\section{Linear Response Calculation}
\label{linres} 

In order to get the low-frequency ac-response of our system it is necessary
to distinguish two contributions to the current. On one hand we have the
current flow $I^{bare}$ of non interacting particles which can be accessed by
a linear response theory. On the other hand we may not neglect the
screening currents $I^{scr}$ due to interactions. 
The low-frequency conductance matrix can be
generally written as
\begin{equation}
\label{Conductance}
G_{\alpha\beta} = \frac{\partial \langle I_{\alpha} \rangle}{\partial V_{\beta}}
= G_{\alpha\beta}^0 - i\omega E_{\alpha\beta},
\end{equation}
where the "emittance" matrix $E$ consists of two parts
$E=E^{bare}+E^{scr}$.

The screening currents may be
calculated quasi-sta\-ti\-cal\-ly solving a Poisson equation self-consistently.
This procedure is described in detail in reference \cite{Buttiker1}.
Here we cite only the result 
\begin{equation}
\label{scremitt} 
E_{\alpha\beta}^{scr} = -\int dr dr' 
g(r,r')\frac{\partial \rho(r)}{\partial V_{\alpha}}
\frac{\partial \rho(r')}{\partial V_{\beta}},
\end{equation}
which is valid in the presence of time reversal symmetry.
The kernel $g(r,r')$ is given by 
$\int dr'' 4\pi \tilde{\Pi}(r,r'') g(r'',r') - \nabla^2_r
g(r,r') = 4\pi \delta(r-r')$.
In a discretized model the Laplace-operator may be
replaced by a capacitance matrix.

To find the bare contribution $E^{bare}$ to the emittance we
proceed as in reference \cite{Pretre1}. We use the current
operator at the normal conducting terminal $\alpha$
\begin{equation}
\label{Current Operator}
\begin{array}{ll}
\hat{I}_{\alpha} = & \frac{1}{h}\sum_{\beta\beta'vww'}q^v \int dE dE' e^{+i(E-E')t/\hbar}\\
\quad\\
                   & \hat{a}_{\beta}^{w\dagger}(E)\hat{a}_{\beta'}^{w'}(E') \times
                     A_{\beta\beta'}^{ww'}(\alpha,v,E,E'),\\
\end{array}
\end{equation}
in a simplified form valid in 
the low-frequency limit.
The full current operator has been given for example in reference \cite{Datta2}.
In Eq. (\ref{Current Operator}) the indices $\alpha,\beta,\beta'$ denote leads (and channels), $v,w,w'$ distinguish
particle and hole states.
The operator $\hat{a}_1^h(E)$ for example creates a hole  
of energy $E$ incident into lead $1$. The matrix elements $A$ are given by
\begin{equation}
\label{Matrix Elements}
A_{\beta\beta'}^{ww'}(\alpha_v,E,E') = 
\delta_{\alpha\beta}\delta_{\alpha\beta'}\delta_{vw}\delta_{vw'}
- \left(s_{\alpha\beta}^{vw}(E)\right)^*s_{\alpha\beta'}^{vw'}(E').
\end{equation}
The ac-conductivity can then be obtained from
\begin{equation}
\label{Linear Response}
g_{\alpha\alpha'} = \frac{1}{\hbar\omega}\int_0^{\infty}d\tau e^{i(\omega+i\delta)\tau}
\langle\left[\hat{I}_{\alpha}(\tau),\hat{I}_{\alpha'}(0)\right]\rangle.
\end{equation}
The evaluation of the commutator is mostly straight forward. As in previous works \cite{Pretre1}
we use the unitarity of the scattering matrix and the thermal occupation of
the reservoirs. As in the case of a purely normal system we are left with
a doubled energy integral. We can evaluate this integral through a path
deformation in the upper complex plane where the scattering matrix is
analytical. In the end we expand the result up to first order in frequency.
The result for the dc-conductance 
\begin{equation}
\label{DC Current}
G_{\alpha\beta}^0 = \frac{1}{h} \int dE \left(-\frac{\partial f}{\partial E}\right)\sum_{\mu\nu}
q^{\mu}q^{\nu}
\left[\delta_{\alpha\beta}\delta_{\mu\nu}
- T_{\alpha\beta}^{\mu\nu}\right]\\
\end{equation}
is identical to the one established 
in the literature \cite{Datta2,Takane1,Beenakker1,Lambert1}.
This serves as a check of our calculation.
In Eq. (\ref{DC Current}), $T_{\alpha\beta}^{\mu\nu} = (s_{\alpha\beta}^{\mu\nu})^*s_{\alpha\beta}^{\mu\nu}$ is
the transmission probability from channel $\beta$ to channel $\alpha$.
The bare emittance can be expressed by global partial densities of states (GPDOS)
\begin{equation}
N(\alpha_{\nu},\beta_{\mu}) = \frac{1}{2\pi} \text{Im} \left(s_{\alpha\beta}^{\nu\mu}\right)^*
\frac{\partial s_{\alpha\beta}^{\nu\mu}}{\partial E},
\end{equation}
and becomes
\begin{equation}
\label{Bare Emittance}
E^{bare}_{\alpha\beta} =  \int dE \left(-\frac{\partial f}{\partial E}\right) \sum_{\nu\mu} q^{\nu}q^{\mu}
N(\alpha_{\nu},\beta_{\mu}).
\end{equation}
This equation shows that the bare emittance may change its sign. This simple
calculation provides only the emittance matrix elements between normal
terminals. A direct calculation of the current at the
superconducting reservoir would involve a self-consistent evaluation
of the pair potential in the superconductor. Its phase ``carries''
the supercurrent. Nevertheless, the missing elements of the emittance matrix
can be reconstructed from the conditions 
\begin{equation}
\label{Conservation}
\sum_{\alpha} E_{\alpha\beta} = \sum_{\beta} E_{\alpha\beta} = 0
\end{equation}
that express current and charge conservation.

\section{Examples}
\label{examples} 

We now present some simple calculations to illustrate how the presence of
a superconducting terminal affects the ac-properties of a mesoscopic
sample. We emphasize that these examples are not designed to model
a realistic sample completely, but should exhibit qualitatively the
main features.

\subsection{Ballistic Wire}
\label{Ballistic Wire}

\begin{figure}[htb]
\begin{center}
\leavevmode
\psfig{file=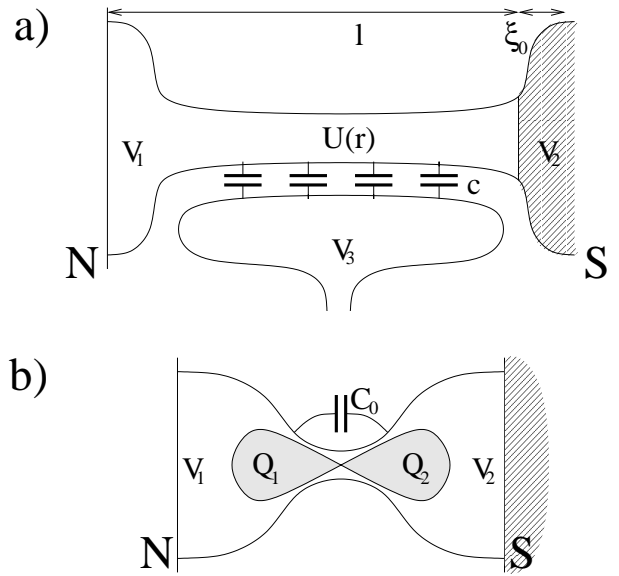,width=8cm}
\vspace{5mm}\\
\caption{Examples:
a) A one channel ballistic wire
is attached to a superconductor. Its length is denoted
by $\ell$, the internal interaction modeled by a
capacitance per unit length $c$ between wire and gate.
The emittance is enhanced by a factor of $4$ compared to
a purely normal wire.
b) The same geometry for a quantum point contact which is
described by an electrostatic dipole of capacitance $C_0$.
In the low transparency limit the ac-conductance is dominated
by the geometric capacitance $C_0$.
In the opposite limit the quantum point contact shows enhanced inductive behaviour.
}
\label{Sketch Ballistic Wire}
\label{Sketch of QPC}
\end{center}
\end{figure}

As a first example we discuss briefly the emittance of a ballistic wire with
one open channel at zero temperature. The results can be easily generalized to more than
one channel. The geometry of the sample is shown in Fig. \ref{Sketch Ballistic Wire}.
The wire is attached to two reservoirs ($1,2$). Reservoir $1$ is always
normal conducting. The second may be normal
conducting, superconducting or completely disconnected from the wire
for comparison. 

The wire is described by its length $\ell$
and the dimensionless parameter $a=4q^2/hv_F$ which
describes the DOS at the Fermi level in the wire.
Note that $a$ is also the local Lindhard function
of Eq. (\ref{Lindhard Function}). 
The interaction in the wire is modeled by a third
gate terminal ($3$). 
It is coupled to the wire by a geometrical capacitance
per unit length $c$ and assumed to be macroscopic.
Thus we can replace the Laplace operator in Eq. (\ref{scremitt}) 
by $-4\pi c$.
For a detailed description
of this system see for example \cite{Blanter1}.

\begin{table} [htb]
\begin{tabular}{l|ccc}
&&&\\
     & normal conducting & superconducting & disconnected\\
&&&\\
\hline
&&&\\
$E_{11}$ & $-\frac{a^2\ell}{4(c+a)}$    & $-a(\ell + \xi_0/2)$                & $\frac{ac\ell}{(c+a)}$ \\
&&&\\
$E_{12}$ & $\frac{a\ell(a+2c)}{4(c+a)}$ & $a(\ell + \xi_0/2)$                           & $0$                          \\
&&&\\
$E_{22}$ & $-\frac{a^2\ell}{4(c+a)}$    &  $-\frac{a^2\ell}{(c+a)}-a\xi_0/2$ & $0$                          \\
&&&\\
\end{tabular}
\caption{Comparison of the emittance elements of a ballistic wire 
connected to two normal reservoirs, a normal and 
a superconducting reservoir and to one normal reservoir only.\label{Emittance Elements}}
\end{table}

As a last parameter we need the coherence length of
the superconductor $\xi_0 = \hbar v_{F}/|\Delta|$. 
We neglect the self-consistency
of its pair potential. 

Table \ref{Emittance Elements} summarizes
the results for the three cases.
The missing elements of the emittance matrix can be
reconstructed from Eq. (\ref{Conservation}).
We add some observations to explain the differences
between the results. The response of the disconnected
wire is purely capacitive, while the open wires
act inductively. In the limit of charge neutrality
$c\ll a$ the inductive emittance of an open wire
$E_{11}$ grows by a factor of four
in the presence of a superconductor. On one hand
the bare emittance is doubled, because an incoming electron
leaving as an Andreev reflected hole stays twice
as long in the wire. On the other hand this effect is
not weakened by a capacitive screened emittance,
because the injectivity from the normal lead into
the wire $\partial\rho/\partial V_1$ is zero.
This leads to another factor of two.
Additionally, the evanescent quasi-particle wave contributes to the bare emittance,
the wire acquires an effective length $\ell+\xi_0/2$
(we use the assumption that the Fermi velocities are
the same on both sides of the NS-interface).

The emittance $E_{13}=-E_{11}-E_{12}$ is always zero in the presence
of a superconductor. The gate and the normal terminal
are only connected via the capacitance. But this capacitance
cannot be charged from the normal side $1$ because the above
mentioned injectivity $\partial\rho/\partial V_1$
is zero. Every injected electron comes back as a hole
that compensates its charge. Therefore, the ac-response to
a bias at the normal end must be zero.
Vice versa the capacitive element $E_{23}$ becomes twice
as big because of a doubled injectivity 
$\partial\rho/\partial V_2$ in the limit of
potential neutrality $c\gg a$.

\subsection{Resonator}
\label{Resonator}

As briefly discussed after Eq. (\ref{Lindhard Function})
we expect the internal charge response to be very
different from a purely normal system, if the states inside
the superconducting gap give the dominant contribution to the Lindhard function.
To further investigate this point
we now consider the charge on a series of
resonances
of Breit-Wigner type coupled to a superconductor.
For analytical simplicity we take the resonances {\it equidistant}. 
A single resonance of this kind is discussed for example in \cite{Beenakker2}.
Its experimental equivalent might be a level
on a quantum dot which is coupled to one normal
and one superconducting lead through tunneling
barriers. However, this situation is normally
treated in a Coulomb blockade picture where
the strongly fluctuating electrostatic potential
is treated as an operator. In our theory the
potential becomes a c-number. Therefore, this model is limited 
to the case of a small charging energy.

The resonances are described by the following
scattering matrix in the normal part
\begin{equation}
s_{N,ij} = \delta_{ij} - i\pi\frac{\sqrt{\Gamma_i\Gamma_j}}{\Delta_L}
\cot\pi\frac{E-qU+i(\Gamma_N+\Gamma_S)/2}{\Delta_L}.
\end{equation}
The coupling width of the resonances to the normal lead
is $\Gamma_N$, to the superconducting lead it is $\Gamma_S$.
The level spacing on the resonator is given by $\Delta_L$.
The above given scattering matrix is only unitary in the
limit $\Gamma_{N/S} \ll \Delta_L$ which means for small
coupling.

\begin{figure}[htb]
\begin{center}
\leavevmode
\psfig{file=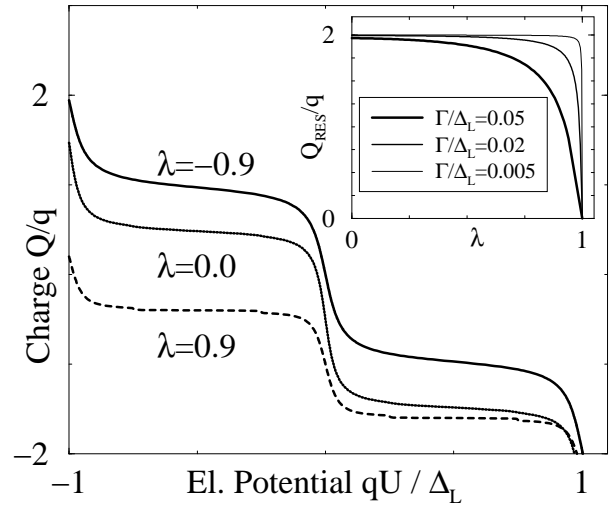,width=8cm}
\caption{
A resonator with level spacing $\Delta_L$ is
attached to a normal and a superconducting wire with
resonance widths $\Gamma_{N/S}$ respectively. Their average
is called $\Gamma = (\Gamma_S+\Gamma_N)/2$, the
parameter $\lambda=(\Gamma_S-\Gamma_N)/(\Gamma_S+\Gamma_N)$
describes their relation. The big plot shows the
charge $Q$ on the resonator as a function of its electrostatic
potential $U$ for different parameters $\lambda$.
The lineshape changes (see Eq. (\ref{Lineshape})), and for
increasing $\lambda$ the charging decreases.
This is illustrated in the inset that shows the charge per resonance
$Q_{RES}$ as a function of $\lambda$.
}
\label{Figure Resonances}
\end{center}
\end{figure}

This scattering matrix is periodic in the energy $E$ and the
integrals in the equation for the charge density (\ref{Charge Density})
diverge. We therefore introduce a cutoff $E_{min}$ by hand that defines
the lowest level on the resonator. This cutoff lies still inside the
energy range of the superconducting gap $|E_{min}| \ll |\Delta|$. This allows
us to use the scattering matrix
\begin{equation}
s_A = 
\left(\begin{array}{cc} s^{pp} & s^{ph}\\ s^{hp} & s^{hh}\end{array}\right) = 
\left(\begin{array}{cc} 0 & -i \\ -i & 0\end{array}\right)
\end{equation}
to describe the Andreev reflection. This example is now constructed 
in such a way that the correction of the Lindhard function due to 
the presence of the superconductor (second term 
in Eq. (\ref{Lindhard Function})) 
becomes very important. 
Fig. \ref{Figure Resonances} shows
a numerical evaluation of the charge on the resonator $Q$ as a function
of the electrostatic potential $U$ on the resonator. 
We describe the coupling by a parameter $\lambda=(\Gamma_S-\Gamma_N)/(\Gamma_S+\Gamma_N)$.
This parameter gets $1$ if the resonance is only connected to the
superconductor and $-1$ in the opposite case.
The inset if Fig. \ref{Figure Resonances} illustrates
the charge stored on one resonance $Q_{RES}$ as a function of $\lambda$. This charge is $2q$ if
the resonator is charged only from the normal side (we assume spin degeneracy, this
is the point where the Coulomb blockade gives a completely different picture). If
the resonator is only connected to the superconductor no charge can be transferred at all.
Below the superconducting gap $|\Delta|$ the states in the resonator are equally
weighted superpositions of particle and hole states that cannot contribute to
the charging of the resonator.

It is worth to note, that the shape of the steps in 
Fig. \ref{Figure Resonances} changes as a function
of $\lambda$. In the two different limits we obtain

\begin{equation}
\label{Lineshape}
\begin{array}{ll}
Q_N = -q\frac{2}{\pi} \text{arctan}\frac{2qU}{\Gamma_N} & \Gamma_N \gg \Gamma_S,\\
\quad\\
Q_S \propto -q \frac{2qU}{\sqrt{(2qU)^2+\Gamma_S^2}}    & \Gamma_S \gg \Gamma_N.
\end{array}
\end{equation}
These results are valid at zero temperature. 
For $kT>\Gamma_{N/S}$ the steps are
smeared out by the Fermi function.

\subsection{Quantum Point Contact}

The low-frequency conductance of a quantum point contact (QPC)
connecting two normal leads has been studied in Ref. \cite{Christen1}.
We adapt this procedure to our situation sketched in Fig. \ref{Sketch of QPC}.
In a first step we only consider one transmission channel.
We assume the QPC is described 
by a symmetric equilibrium potential. At equilibrium the only 
asymmetry stems 
from the presence of the superconducting lead. 
Polarization of the QPC due to an applied voltage leads 
to a dipole. (Charging vis-avis the gates is neglected. See however Ref. 
\cite{Christen2}). 
The size of this dipole 
is described by one single capacitance $C_0$. Furthermore, we
limit ourselves to a semiclassical treatment which essentially
means that the confining potential is sufficiently expanded in space.
In this limit the second and third part of Eq.
(\ref{Lindhard Function}) are negligible.
As a second parameter we need the total density of states at the Fermi level
(over a region in which the charge is not screened fully), 
when the system is entirely normal
$D^N = q^2N_F$. In addition, scattering at the QPC 
is characterized by its transmission probability $T$ and its reflection
probability $R = 1-T$. As shown in \cite{Christen1} the
electrochemical capacitance and the emittance in the purely
normal system are
\begin{equation}
\label{QPC Normal Result}
C^N = R\frac{C_0 D^N}{4C_0 + D^N}, \quad E^N = RC^N - \frac{D^N}{4}T^2.
\end{equation}
This result uses the fact that the semiclassical
injectivities may be written as
\begin{equation}
\frac{\partial Q^N_1}{\partial V_1} = \frac{D^N}{4}\left(1+R\right), \quad
\frac{\partial Q^N_2}{\partial V_1} = \frac{D^N}{4}\left(1-R\right).
\end{equation}
For example, the response $\partial Q^N_1/\partial V_1$ 
originates from all
the right going electrons in region $1$ plus the left going ones that have
been reflected at the barrier.

It is clear that this picture will change drastically in the
presence of a superconductor. We denote by $R_N=4R/(1+R)^2$
the probability that an electron is scattered back as
an electron. The probability for Andreev reflection we
call $R_A=1-R_N=T^2/(1+R)^2$. The dc-conductance is
of course $4R_Aq^2/h$. The injectivities now turn out to
be
\begin{equation}
\frac{\partial Q^S_1}{\partial V_1} = \frac{D^N}{2}R_N ,\quad
\frac{\partial Q^S_2}{\partial V_1} = 0 ,\, 
\end{equation}
\begin{equation}
\frac{\partial Q^S_1}{\partial V_2} = \frac{D^N}{2}R_A , \quad
\frac{\partial Q^S_2}{\partial V_2} = \frac{D^N}{2}.
\end{equation}
For example in $\partial Q^S_1/\partial V_1$, we recognize
that only the electrons that return as electrons contribute
to the injectivity. We see also that the normal terminal
cannot inject charge into the right side of the QPC which
is also clear from intuition.

Now we use these ingredients to find the capacitance and
emittance of the whole QPC. For simplicity, we cite the results without
the length renormalization due to a finite $\xi_0$ in the
superconductor. We find 
\begin{equation}
\label{QPC Super Capacitance}
C^S = R_N \frac{C_0 D^N}{4C_0 + D^N} , 
\end{equation}
\begin{equation}
\label{QPC Super Emittance}
E^S = R_N C^S - \frac{T(1-4R-R^2)}{(1+R)^4} D^N.
\end{equation}

In the low transparency limit ($R\simeq 1$) the result
is the same as for the purely normal conducting system (\ref{QPC Normal Result}).
In the high transparency limit ($R\simeq 0$) we recover the inductive behaviour
of example \ref{Ballistic Wire}. Again the emittance is increased by a
factor of four in comparison to the result (\ref{QPC Normal Result}).

\begin{figure}[htb]
\begin{center}
\leavevmode
\psfig{file=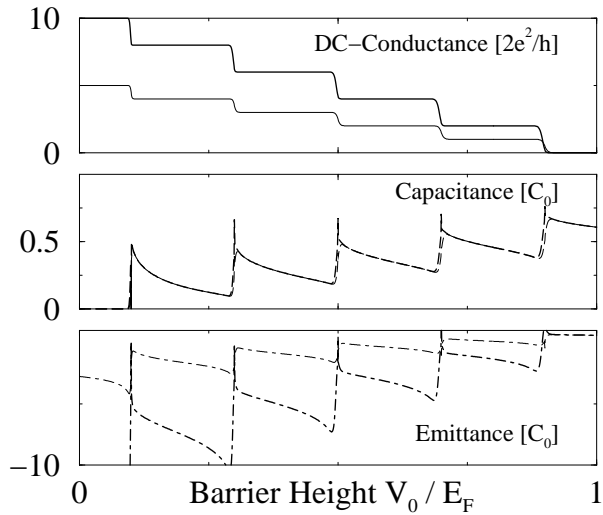,width=8cm}
\caption{Comparison between transport properties of a multichannel
quantum point contact either attached to a normal lead (narrow line) 
or a superconducting lead (broad line). In the presence of the
superconductor, its dc-conductance is doubled
and the low-frequency ac-response is enhanced by a factor of four
in the transparent limit. The curves are calculated in a
WKB-approximation. The used parameters are specified
in the text.
}
\label{QPC Graph}
\end{center}
\end{figure}

Fig. \ref{QPC Graph} shows a qualitative comparison
of the conductance, capacitance and emittance of a multichannel QPC
in the two geometries. We use a capacitance of $C_0=1\text{fF}$ and
a potential $U(x)=\text{max}\{V_0(\lambda^2-x^2),0\}$ where
$\lambda=500\text{nm}$. The constriction in y-direction allows
up to five open channels with equidistant spacing through the contact.

What are the restrictions of the results obtained for our simple model
system?
The assumption that the NS-interface is a perfect Andreev mirror seems to play
the most important role. In this case we may neglect the capacitance of
the NS-interface. If such a capacitance 
would be present it would 
decrease the inductive behaviour at high transparency.

\section{Conclusions}
\label{conclusion} 

In this work we have extended the ac-response 
theory of normal mesoscopic conductors 
to hybrid normal and superconducting structures. 
This requires an investigation of screening and
a discussion of the charge density response
to external lead voltages in the presence of 
Andreev scattering.
Global gauge invariance is valid also for the hybrid structures investigated here. 
This leads necessarily to the existence of an {\it injectivity} of the 
superconductor into the normal part of the structure. The charge-injectivity of the 
superconductor compensates the suppression of the charge-injectivity from a normal contact. 

Screening in hybrid structures 
is up to small corrections the same as in normal conductors. 
Nevertheless, the ac-response of hybrid structures exhibits marked differences 
from that of a purely normal system. For a ballistic wire 
at one end connected to a superconducting reservoir, the 
emittance is four times as large as that of a purely normal wire. 
Furthermore, the displacement current induced into a nearby gate in 
response to an oscillating voltage at 
the normal contact (described by an off-diagonal capacitance element) 
is highly suppressed compared to the purely normal structure. 
For a resonant structure, the charge response to an internal variation
of the electrostatic potential can be reduced due to the superposition
of particle and hole states inside the cavity.
A quantum point contact attached to a superconductor shows the same
capacitive behavior as its normal conducting analog in the limit
of small transmission. For high transmission the emittance is enhanced
as in the case of a ballistic wire.

For the ac-conductance problem screening is necessary if we want to find 
a response that depends only on voltage differences and which conserves 
current. We have focused on geometries with a 
single NS-interface but similar considerations should apply if we deal 
with SNS-structures or more complicated geometries. Electrical 
self-consistency is relevant not only for dynamic problems but also 
if we are interested in non-linear transport or even just in 
the gate voltage dependence of stationary transport quantities. 
Therefore, the considerations presented 
should be useful for a wide range 
of geometries and for the investigation of many different physical 
problems.

\section*{Acknowledgements}
We thank Wolfgang Belzig for an important discussion. 
This work was supported by the Swiss National Science Foundation
and by the RTN network on "Nanoscale Dynamics, Coherence and 
Computation".

\appendix

\section{Charge Density}
\label{Anhang A}

This is a short sketch of the derivation of for\-mu\-la (\ref{Charge Density}).
We ex\-press the ex\-pec\-ta\-tion value of the charge den\-si\-ty operator
with help of the normalized solutions 
$(w_{\beta}^{\mu}(r,E))^{\dagger} =((u_{\beta}^{\mu}(r,E))^*,(v_{\beta}^{\mu}(r,E))^*)$
of the Bogoliubov-de Gennes equation \cite{Datta1}
\begin{equation}
\label{Expectation Value}
\langle \hat{\rho}\rangle =  q\sum_{\beta\mu}\int^{+\infty}_{-\infty} dE 
(w_{\beta}^{\mu})^{\dagger}
\left(\begin{array}{cc} f_{\beta}^{\mu} & \\ & 1-f_{\beta}^{\mu}\end{array}\right)
w_{\beta}^{\mu}.
\end{equation}
The solutions of the Bogoliubov-de Gennes equation are scattering
states in our case. We include the usual prefactors containing the group
velocities in the normalization factors of the wavefunctions. 
Their mean occupation number can be expressed by the
Fermi functions of the reservoirs $f_{\beta}^{p/h}(E) = f(E-q^{p/h}V_{\beta})$ depending on whether they
describe an incoming particle $p$ or hole $h$. $q^{p/h} = \pm q$ denotes their quasi-particle charge.

The starting point of the calculation is the Bogoliubov-de Gennes equation (\ref{Bogoliubov-de Gennes}).
For the moment we allow the electrostatic potentials $U^{p/h}$ to be complex. The continuity
equation for the quasi-particle current $j_{\alpha\beta}^{\mu}$ then reads
\begin{equation}
\label{Continuity Equation}
\nabla j_{\beta}^{\mu} = \frac{2q}{\hbar}
\left[\text{Im}(U^p)\left|u_{\beta}^{\mu}\right|^2
 + \text{Im}(U^h)\left|v_{\beta}^{\mu}\right|^2 \right].
\end{equation}
The complex potentials generate source terms on the right side of Eq. 
(\ref{Continuity Equation}). As a next step we integrate this equation over
the volume of the scatterer. To this end we choose the potentials to vary like
$U^{p/h} = U^{p/h}_0 + i\Gamma^{p/h}\delta(x-x_0)$. We then get for the
current
\begin{equation}
I_{\beta,out}^{\mu} - I_{\beta,in}^{\mu} = \frac{2q}{\hbar}
\left[\Gamma^p \left|u_{\beta}^{\mu}(x_0)\right|^2
+ \Gamma^h \left|v_{\beta}^{\mu}(x_0)\right|^2 \right].
\end{equation}
In this equation we call $I_{in/out}$ the total current flow
into and out of the scattering region. Their difference is not
zero because of the source term in Eq. (\ref{Continuity Equation}).
The ratio of both quantities can be  expressed by the scattering matrix $s$
\begin{equation}
\frac{I_{\beta,out}^{\mu}}{I_{\beta,in}^{\mu}} =
\sum_{\alpha\nu} \left(s_{\alpha\beta}^{\nu\mu}\right)^* s_{\alpha\beta}^{\nu\mu}.
\end{equation}
The scattering matrix is a functional of the small complex variation
$\Gamma_{p/h}$ and thus can be expanded up to first order. To evaluate the
incoming current $I_{in}$ we use the normalization of the wave functions
$w_{\beta}^{\mu}$ and get $I_{\beta,in}^{\mu} = 1/h$. 
Finally we manage
to express the square of the wave functions by the LPDOS given in definition
(\ref{Partial Densities})
\begin{equation}
\label{Wave Squares1}
\sum_{\alpha\nu} \nu(\alpha_{\nu},r_{p},\beta_{\mu}) = \left|u_{\beta}^{\mu}(r)\right|^2,
\end{equation}
\begin{equation}
\label{Wave Squares2}
\sum_{\alpha\nu} \nu(\alpha_{\nu},r_{h},\beta_{\mu}) =
\left|v_{\beta}^{\mu}(r)\right|^2.
\end{equation}
These quantities can be inserted into Eq. (\ref{Expectation Value}) to get the
final result (\ref{Charge Density}) given at the beginning of the article.

\section{Lindhard Function}
\label{Anhang B}

In this appendix we explain the derivation of the local Lindhard function (LLF)
given in Eq.
(\ref{Lindhard Function}). The nonlocal Lindhard function (NLF) is given by
$\tilde{\Pi}(r,r')=-\delta \rho(r) / \delta U(r')$. We define functional
potential derivatives of the LPDOS as follows
\begin{equation}
\chi_{\alpha\beta}^{\mu\nu}(r_{\kappa},r'_{\lambda}) = 
\frac{\delta \nu(\alpha_{\nu},r_{\kappa},\beta_{\mu})}{\delta U^{\lambda}(r')}.
\end{equation}
Using this definition and Eq. (\ref{Charge Density}) we can write the NLF as 
\begin{equation}
\label{Nonlocal Lindhard Function}
\begin{array}{ll}
\tilde{\Pi} = & -q\int_{-\infty}^{\infty} dE \sum_{\alpha\beta\mu\nu\lambda}\\
\quad\\
& \left(
f(E) \chi_{\alpha\beta}^{\mu\nu}(r_p,r'_{\lambda})
 + f(-E) \chi_{\alpha\beta}^{\mu\nu}(r_h,r'_{\lambda})
\right).
\end{array}
\end{equation}
The NLF is not a Fermi surface quantity but depends on 
all energies within the conduction band. 
The LLF is a good approximation if the electrostatic potential $U$ varies
slowly on the scale of the Fermi wavelength. Under these circumstances the spatial integration appearing
for example in Eq. (\ref{Charge Response}) can be simplified
\begin{equation}
\int dr' \tilde{\Pi}(r,r') \delta U(r') = \delta U(r) \int dr' \tilde{\Pi}(r,r') = \delta U(r) \Pi(r).
\end{equation}
To get the LLF we must therefore integrate the NLF over its second spatial variable $r'$.
Because of particle-hole symmetry it is sufficient to keep the first part of
Eq. (\ref{Nonlocal Lindhard Function}). We may thus write the LLF in the following way
\begin{equation}
\label{Splitted Integral}
\Pi = -2q\left( \int\limits_{-\infty}^{-E_c} + \int\limits_{-E_c}^{+\infty}\right)dE f(E) \sum_{\alpha\beta\nu\mu\lambda}
\int dr' \chi_{\alpha\beta}^{\mu\nu}(r_p,r'_{\lambda}), 
\end{equation}
where we used an energy scale $E_c$ with the property $|\Delta|\ll E_c \ll E_F$. Below $-E_c$ we
may neglect any Andreev reflection. 
In this case there is no coupling between particle and hole states and all 'crossed' LPDOS
and the functional derivative $\delta / \delta U^h(r')$ vanish. 
We may furthermore use 
\begin{equation}
\label{Lindhard Relation}
\frac{\partial}{\partial E_p} = -\int dr'\frac{\delta}{q\delta U_p(r')}, \quad
\frac{\partial}{\partial E_h} = +\int dr'\frac{\delta}{q\delta U_h(r')}.
\end{equation}
which holds in WKB-approximation 
and $\partial/\partial E^p = d/dE$
to get
\begin{equation}
\label{Below Gap}
\int dr' \chi_{\alpha\beta}^{\mu\nu}(r_p,r'_{\lambda}) \simeq
\delta_{p\nu}\delta_{p\mu}\delta_{p\lambda} \left(-\frac{d}{dE}\right) \nu(\alpha_p,r_p,\beta_p).
\end{equation}
The second integral may similarly be simplified using 
Eq. (\ref{Lindhard Relation}). Applying 
partial integration with 
respect to the energy we get the result 
presented in Eq. (\ref{Lindhard Function}).

\end{document}